\documentclass[reprint,
 amsmath,amssymb,
 aps
]{revtex4-2}
\usepackage{physics}
\usepackage{graphicx}
\usepackage{dcolumn}
\usepackage{bm}
\usepackage{caption}
\usepackage{bbold}
\usepackage{subcaption}
\usepackage{xcolor}
\usepackage{tikz}
\usetikzlibrary{decorations.pathmorphing}
\tikzset{zigzag/.style={decorate, decoration=zigzag}}

\definecolor{darkgreen}{HTML}{006622}
\newcommand{\comment}[1]{}
\begin{document}

\preprint{APS/123-QED}

\title{Signatures of the black hole quantum atmosphere in nonlocal correlations}% Force line breaks with \\

\author{Adam Z. Kaczmarek}\email{a.kaczmarek@doktorant.ujd.edu.pl}
\author{Dominik Szcz{\c{e}}{\'s}niak}
\affiliation{Department of Theoretical Physics, Faculty of Science and Technology, Jan D{\l}ugosz University in Cz{\c{e}}stochowa, 13/15 Armii Krajowej Ave., 42200 Cz{\c{e}}stochowa, Poland}
\date{\today}% It is always \today, today,

\begin{abstract}
Recently, it was suggested that the Hawking radiation may originate not at the event horizon but in the quantum region outside of it, known as the {\it quantum atmosphere}. The present study attempts to explore this argument further by assessing its role in shaping quantum correlations near a black hole. Herein, these are conveniently captured within the geometric measure of nonlocality, termed as the measurement-induced nonlocality, and found to exhibit signatures of the atmosphere. In particular, a notable loss of correlations is observed well outside the event horizon, coinciding with the peak of particles radiation in the atmosphere region. Still, the correlations are shown to be always finite therein and to continuously scale with not only the radiation temperature but also with the horizon size. Hence, some characteristics of the atmosphere appears to be detectable at the quantum correlations level, providing novel insight and means to help verify the concept of interest.
\end{abstract}

\maketitle

%%%%%%%%%%%%%%%%%%%%%%%%%%%%%%%%%%%%%%%%%%%%%%%%%%%%%%%%%%%%%%%%%%%%%%%%%%%%%%%%%%%%%%%%%%%%%%
\section{Introduction}
%%%%%%%%%%%%%%%%%%%%%%%%%%%%%%%%%%%%%%%%%%%%%%%%%%%%%%%%%%%%%%%%%%%%%%%%%%%%%%%%%%%%%%%%%%%%%%

In the 1970s, the studies of quantum fields on a curved background have led to the famous results by Hawking \cite{hawking1975}. In his pioneering work, Hawking has shown that black holes evaporate due to the particle creation. Although providing groundbreaking insight into the behavior of gravity at the quantum level, the Hawking radiation also raised some questions. One of the biggest and still unresolved challenges in this regard is the information paradox, stating that the evaporating black hole evolves from its initial to final state in contradiction to the unitary time evolution of quantum mechanics \cite{hawking1976}. In the pursuit to unravel this inconstancy, it is crucial to ask, among other questions, where exactly the radiation quanta originate? The original investigations by Hawking and Unruh suggested that it arises from the quantum excitations at the effective distance ($r$) very near the event horizon ($\Delta r=r-r_{H} \ll r_{H}$, where $r_{H}$ is the Hawking radius) \cite{hawking1975, unruh1977}. However, this viewpoint was recently contested by Giddings, advocating for the notion of the source region ($r_{A}$) well outside the event horizon ($\Delta r=r_{A}-r_{H} \sim r_{H}$), known as the {\it quantum atmosphere} \cite{giddings2016}.

The above claim by Giddings was initially supported by the estimates of the effective size of an emitting body, as based on the Stefan-Boltzmann law, and the simultaneous calculations of the Hawking quanta wavelength \cite{giddings2016}. Later on, the follow-up studies on the stress energy tensor \cite{dey2017, dey2019} the heuristic gravitational Schwinger effect argument \cite{dey2017, ong2020} or the quantum correlations across the event horizon \cite{balbinot2022} only backed up this idea. Interestingly, it was also shown that the firewall phenomenon \cite{almheiri2013} is somewhat compatible with the quantum atmosphere \cite{kim2017} and that the location of the latter can be well estimated based on the thermal behavior of the radiation \cite{eune2019}.

This intriguing concept of quantum atmosphere naturally calls for an even deeper insight, not only in terms of the Hawking radiation origin but also the resulting alternation of a black hole surrounding. An intuitive step in this direction is to perform measure of some kind in the background of a black hole. Due to the intrinsically quantum character of the Hawking radiation, the inspection of quantum correlations in the vicinity of the event horizon appears as a suitable approach for this task. In particular, it is expected here that the distinct structure of Hawking quanta should have profound impact on a quantum system, allowing to trace its behaviour under the new setting and to observe potential signatures of the atmosphere.

In the present study, we explore quantum atmosphere from such a new perspective. This is done by evoking the phenomenon of nonlocality, a fundamental and potentially frame-independent property of any quantum system or reality in general \cite{brunner2014}. In a nutshell, the setup of two entangled parties is considered here to be located near a black hole, allowing to capture thermal characteristics of its atmosphere in analogy to the earlier studies on quantum systems in the Schwarzschild or the Garfinkle-Horowitz-Strominger space-times \cite{he2016, kaczmarek2023}. In this context, the nonlocality itself is quantified within the so-called measurement-induced nonlocality (MIN), a genuine correlation measure between parts of the composite system \cite{luo2011}. The MIN is well-suited for such considerations since it describes correlations in a broad manner, going beyond the Bell theorem and allowing for nonlocality without entanglement or the nonlocality without quantumness \cite{hu2012}. This is to say, the assumed approach allows to conveniently probe the atmosphere region and to interpret its role in shaping quantum correlations near a black hole. As a result, the corresponding analysis is expected to provide vital contribution to the field aimed at comprehending the nature of the Hawking radiation via the correlation measures \cite{pan2008, farahi2014, he2016, hu2018, wu2022, kaczmarek2023}.

The present study is organized as follows: in section \ref{methods1} we present description of the quantum systems of interest in the Schwarzschild space-time, next in section \ref{methods2} we define our setup and analytically outline the measurement-induced nonlocality in the presence of the quantum atmosphere, finally section \ref{results} gives our predictions on the thermal evolution of quantum correlations in the Hartle-Hawking vacuum. To this end, the obtained results are summarized by some pertinent conclusions.

%%%%%%%%%%%%%%%%%%%%%%%%%%%%%%%%%%%%%%%%%%%%%%%%%%%%%%%%%%%%%%%%%%%%%%%%%%%%%%%%%%%%%%%%%%%%%%
\section{Dirac fields in the Schwarzschild space-time}
\label{methods1}
%%%%%%%%%%%%%%%%%%%%%%%%%%%%%%%%%%%%%%%%%%%%%%%%%%%%%%%%%%%%%%%%%%%%%%%%%%%%%%%%%%%%%%%%%%%%%%

In the present analysis, we choose the initial state to be of the fermionic type. This allows us to be on the same footing with other recent studies on quantum correlations in the relativistic setting, which frequently consider Dirac particles \cite{hu2018, huang2019, li2022, wu2022, kaczmarek2023}. In this regard, it is instructive to note also that the fermionic modes are more resistant towards the Hawking radiation than the bosonic ones and have non-zero value even for the radiation temperature approaching infinity \cite{he2016, hu2018, kaczmarek2023}.

In the context of the above, the Dirac equation for the curved space-time is first considered:
\begin{align}
    (i\gamma^a e^\mu_aD_\mu-m)\psi=0,
    \label{eq1}
\end{align}
where $D_\mu=\partial_\mu-\frac{i}{4}\omega^{ab}_\mu \sigma_{a b}$, $\sigma_{ab}=\frac{i}{2}\{\gamma_a,\gamma_b\}$, $e^\mu_a$ is {\it vierbein} and $\omega^{ab}_\mu$ denotes the spin connection. The positive frequency solutions of Eq. ({\ref{eq1}) correspond to the regions $I$ and $II$ {\it i.e.} outside and inside the event horizon ($r=r_h$), respectively. In order to obtain a complete basis for the analytic modes with the positive energy, the Kruskal coordinates are utilized to perform analytical continuation in accordance to the Damour-Ruffini method \cite{damour1976, dey2017}. The resulting Dirac fields are expanded in the appropriate Kruskal basis, as follows:
\begin{align}\nonumber
    \Psi&=\sum_i d\textbf{k}\frac{1}{\sqrt{2\cosh(\pi \omega_i/\kappa)}}\\
&\times\Big[ c^I_\textbf{k}\zeta^{I+}_\textbf{k}+c^{II}_\textbf{k}\zeta^{II+}_\textbf{k} +d^{I\dagger}_\textbf{k}\zeta^{I_-}_\textbf{k}+d^{II\dagger}_\textbf{k}\zeta^{II_-}_\textbf{k}\Big],
    \label{eq2}
\end{align}
where $c_\textbf{k}$ and $d_\textbf{k}^\dagger$ are subsequently the creation and annihilation operators applied to the Kruskal vacuum \cite{pan2008}. Next, by using the Bogoliubov transformation, it is possible to establish the relation between operators in a black hole and the Kruskal space-times \cite{ge2008}. In particular, the vacuum and excited states of the black hole coordinates correspond to the Kruskal two-mode squeezed states as:
\begin{align}  
    \nonumber &\ket{0_{\textbf{k}}}^+=\alpha\ket{0_\textbf{k}}_I^+\ket{0_{-\textbf{k}}}^-_{II}+\beta \ket{1_\textbf{k}}_I^+\ket{1_{-\textbf{k}}}^-_{II},\\
    &\ket{1_{\textbf{k}}}^+=\ket{1}^+_{I}\ket{0_{-\textbf{k}}}_{II}^-,
    \label{eq3}
\end{align}
with the following Bogoliubov coefficients \cite{lanzagorta2014}:
\begin{align}
    \alpha=\frac{1}{(e^{-\omega_i /T}+1)^{1/2}},\;\;\; \beta=\frac{1}{(e^{\omega_i /T}+1)^{1/2}},
     \label{eq4}
\end{align}
where $T$ denotes Hawking temperature of the emitted radiation. The above reasoning can next directly link to the Giddings argument \cite{giddings2016}, by considering a dimensionally reduced Schwarzschild black hole of line element $ds^2=-f(r)dt^2 + f(r)^{-1}dr^2$. 

\begin{figure*}
    \centering
    \includegraphics[scale=0.85]{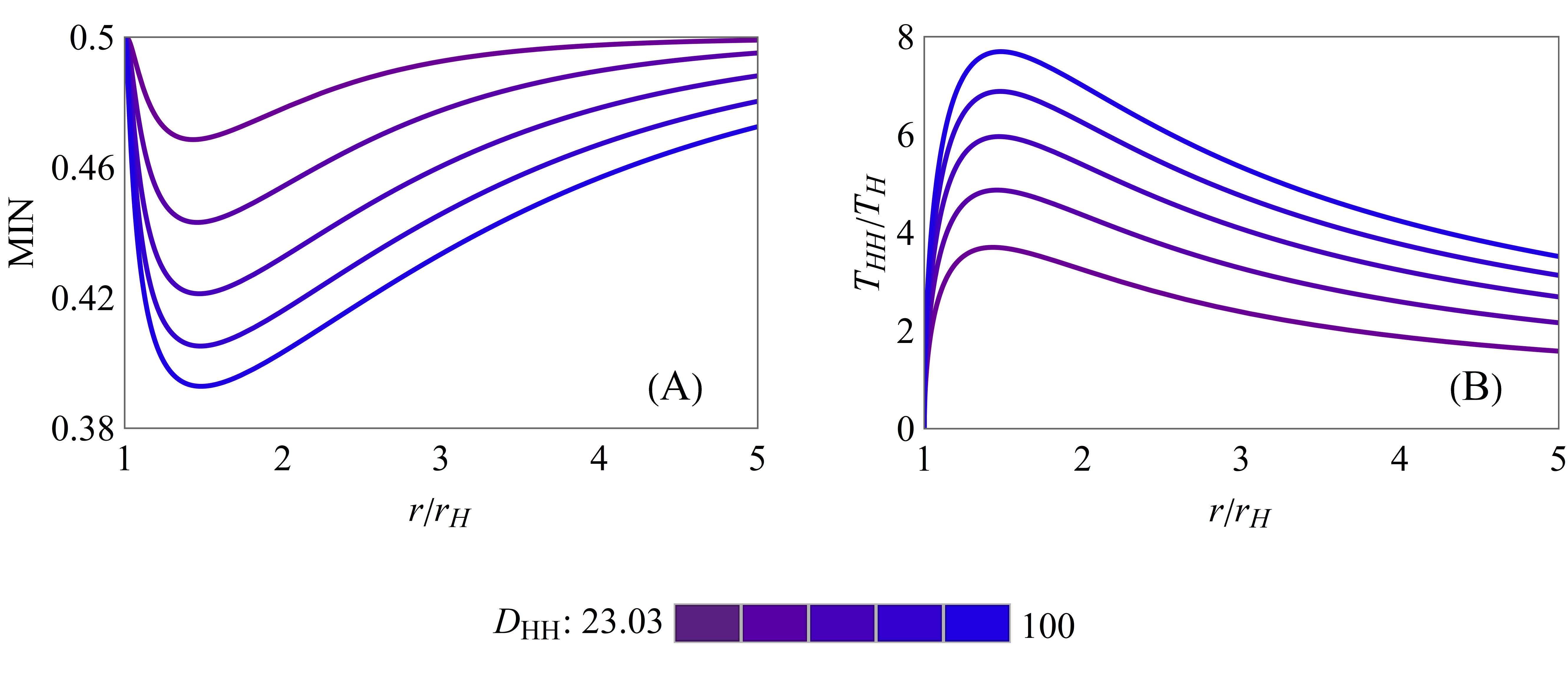}
    \caption{(A) The measurement-induced nonlocality of the physically accessible quantum correlations ($\text{MIN}(\rho_{AB_I})$) and (B) the normalized local temperature in the Hartle-Hawking vacuum ($T_{HH}/T_{H}$) as a function of the normalized distance ($r/r_{H}$) for the selected values of the so-called Hartle-Hawking constant ($D_{HH}$).}
    \label{r1}
\end{figure*}

\section{Measurement-induced nonlocality in the Hartle-Hawking vacuum}
\label{methods2}

To encompass how the origin of the Hawking quanta affects quantum systems in the presence of the quantum atmosphere, the slightly modified version of the well-established probing scenarios \cite{alsing2006, ge2008, pan2008} is employed. In details, it is assumed that Alice and Bob share maximally entangled Bell state of the form:
\begin{align}
    \rho_{AB}=\ket{\phi^+}\bra{\phi^+} \;\;\;\text{for}\;\;\; \ket{\phi^+}=\frac{1}{\sqrt{2}}\big(\ket{00}+\ket{11}\big),
    \label{eq5}
\end{align}
while being equipped with the detectors sensitive only to modes $A$ and $B$, respectively. After that, Alice remains stationary at the asymptotically flat region of the space-time. On the other hand, Bob, after the initial free-fall, starts to hover at the distance $r$ from the black hole centre. In this manner, at some radius $r$ the detector $B$ is expected to experience the outgoing radiation in the quantum atmosphere region, which is extended beyond the radius $r_h$ of an event horizon. To describe what Bob perceives, the corresponding vacuum and excited states are being constructed in the Kruskal frame by using the appropriate Bogoliubov transformation between operators, as described by Eqs. (\ref{eq3}) and (\ref{eq4}) \cite{ge2008, pan2008}. As a consequence, the $\rho_{AB}$ state can be now represented in a new basis, with the Alice and Bob being under the influence of the radiation quanta. In this regard, the density matrix $\rho_{A B_I B_{II}}$ is obtained by expanding Bob mode into the Kruskal one ($B \rightarrow B_I B_{II}$). Since the region $II$ is physically inaccessible, it is necessary to perform the trace over the region $II$ {\it i.e.} $\rho_{A B_I}= \Tr_{B_{II}}{(\rho_{A B_I B_{II}})}$. In this way the density matrix, describing physically accessible correlations with the $A$ and $B$ confined to the physical region $I$, can be obtained \cite{ge2008, he2016, harikrishnan2022}. We remark that since this is a {\it{gedanken}} setup, the present analysis can be considered to be complementary to the recent work regarding quantum correlations across the event horizon \cite{balbinot2022}.

In the outlined framework, the MIN for the bipartite quantum state $\rho$, shared by the $A$ and $B$ parts, is defined to be \cite{luo2011, hu2018}:
\begin{align}
    \text{MIN}(\rho)=\text{Max}_{\Pi^A}\Big|\Big| \rho - \Pi ^A (\rho)\Big|\Big|^2.
\end{align}
In the case of Eq. (\ref{eq5}), the following MIN form for the physically accessible quantum correlations can be obtained by using the Bloch decomposition \cite{tian2013, he2016, kaczmarek2023}:
\begin{align}
    \text{MIN}(\rho_{AB_I})=\frac{1}{2(1+e^{-\omega /T})}. 
    \label{eq7}
\end{align}
We note, that since our goal is to characterize possible signatures of the quantum atmosphere, the local temperature ($T$) should take the Hartle-Hawking ($T_{HH}$) form \cite{eune2019}:
\begin{align}\nonumber
    &T_{HH}=T_H \sqrt{1-\frac{r_h}{r}}\\&\sqrt{1+2\frac{r_h}{r}+\big(\frac{r_h}{r}\big)^2\Big(9+D_{HH}+36 \ln(\frac{r_h}{r})\Big)},
    \label{eq8}
\end{align}
with $T_H=\frac{1}{4\pi r_h}$. In Eq. (\ref{eq8}), the $D_{HH}$ is the undetermined constant of the stress tensor for the Hartle-Hawking vacuum, termed here as the Hartle-Hawking constant for simplicity. According to \cite{eune2019}, its value cannot be fixed by the Hartle-Hawking boundary conditions. However, it is known that for $D_{HH}\geq D_C\approx 23.03$ the temperature is positive and decreases after reaching peak at $r_c \approx 1.43 r_h$. Note that this local temperature is vanishing at the horizon $r_h$ and approaches Hawking temperature at the infinity ($r \rightarrow \infty$). In addition, the $T_{HH}=0$ result agrees with the lack of the influx/flux of the particle radiation for the horizon in thermal equilibrium \cite{eune2019}. At the same time, $T_{HH}$ peaks at the macroscopic distance from the event horizon where the main excitations occur.

%%%%%%%%%%%%%%%%%%%%%%%%%%%%%%%%%%%%%%%%%%%%%%%%%%%%%%%%%%%%%%%%%%%%%%%%%%%%%%%%%%%%%%%%%%%%%%
\section{Thermal evolution of quantum correlations}
\label{results}
%%%%%%%%%%%%%%%%%%%%%%%%%%%%%%%%%%%%%%%%%%%%%%%%%%%%%%%%%%%%%%%%%%%%%%%%%%%%%%%%%%%%%%%%%%%%%%

\begin{figure*}
    \centering
    \includegraphics[scale=0.8
    ]{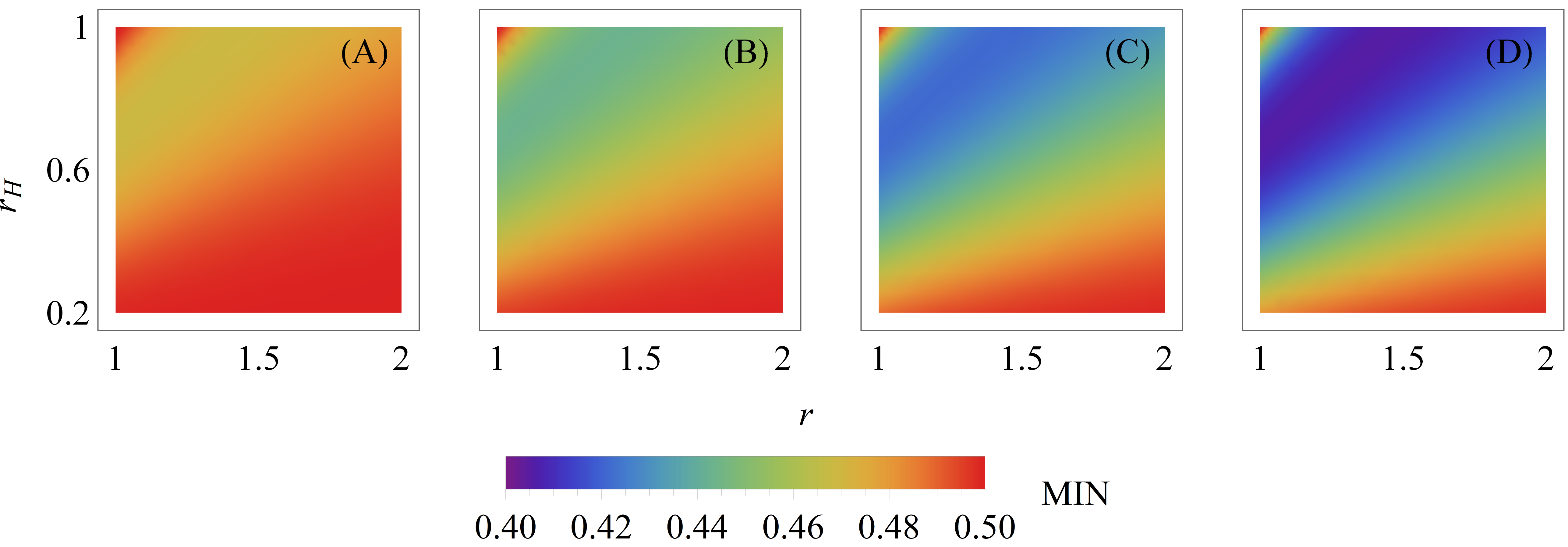 }
    \centering
    \includegraphics[scale=0.8
    ]{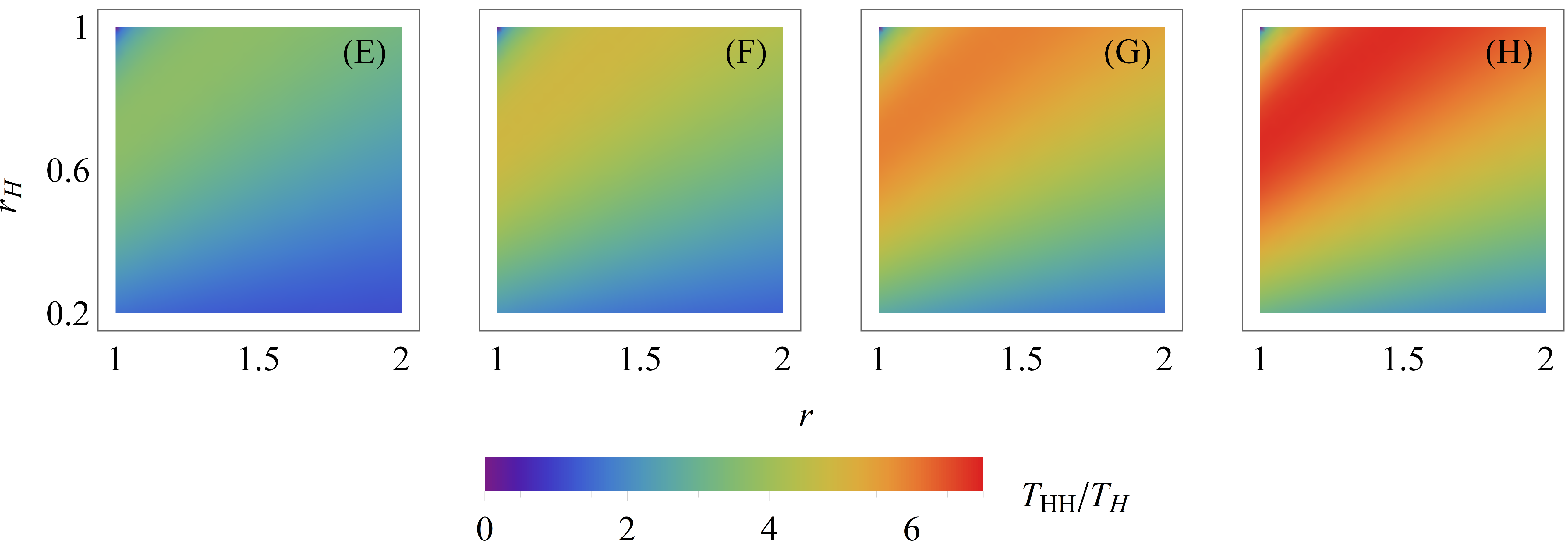}
    \caption{(A)-(D) The measurement-induced nonlocality of the physically accessible quantum correlations ($\text{MIN}(\rho_{AB_I})$) and (E)-(H) the normalized local temperature in the Hartle-Hawking vacuum ($T_{HH}/T_{H}$) on a plane defined by the distance from a black hole ($r$) and the event horizon size ($r_{H}$). The results are depicted for the selected values of the so-called Hartle-Hawking constant ($D_{HH} \in \{23.03,40,60,80\}$, in order from left to the right-hand side).}
    \label{r2}
\end{figure*}

In Fig. \ref{r1} (A), the behaviour of the $\text{MIN}(\rho_{AB_{I}})$ measure, as a function of the normalized distance ($r/r_{H}$), is presented for the selected values of the $D_{HH}$ constant. Note that only $D_{HH}\geq D_C\approx 23.03$ is considered to avoid physically irrelevant solutions, yielding imaginary or inverse distance-dependent temperature \cite{eune2019}. Although the full physical meaning of the $D_{HH}$ is still unknown, the obtained results show that the $\text{MIN}(\rho_{AB_{I}})$ measure behaves qualitatively the same for each of the considered parameter values. In details, it exhibits continuous character showing that the quantum correlations of interest initially decrease but later return to its maximum value at $r > 4 r_{H}$, as the distance from the event horizon increases. It is instructive to note that the $\text{MIN}(\rho_{AB_{I}})$ minimum is located always at $r \approx 1.43 r_{H}$, regardless of the assumed $D_{HH}$ level. The latter is responsible only for the absolute value of the $\text{MIN}(\rho_{AB_{I}})$ measure and the width of the correlation loss region. This is to say, the higher the $D_{HH}$ value the more visible the correlation drop is. Deeper inspection allows relating this loss to the peak of the Hawking quanta radiation well outside the event horizon. In Fig. \ref{r1} (B), the distance-dependent character of the normalized local temperature ($T_{HH}/T_{H}$) is presented for convenience. The depicted results clearly show that the aforementioned peak is indeed located around $r \approx 1.43 r_{H}$, in correspondence to the correlation loss region. Similarly, it is possible to trace back the destructive role of the $D_{HH}$ parameter on the quantum correlations and directly associate it with the local temperature (refer to Fig. \ref{r1} (B)).

To verify if the above claims are true for different horizon radii, the $\text{MIN}(\rho_{AB_{I}})$ measure is examined further by analyzing its dependence on the $r$ and $r_{H}$ distances separately. In Figs. \ref{r2} (A)-(D) the density plots of the $\text{MIN}(\rho_{AB_{I}})$ are given on a plane defined by these two parameters. Once again several values of the $D_{HH}$ constant are chosen to enhance some of the effects ($D_{HH} \in \{23.03,40,60,80\}$). Similarly to results presented in Fig. \ref{r1} (A), the $\text{MIN}(\rho_{AB_{I}})$ shows minimum just outside the event horizon, which aligns with the previously observed ratio $r/r_h \sim 1.43$. Moreover, the destructive influence of the $D_{HH}$ parameter is also confirmed. However, the presented results allows to observe additional aspects of the discussed quantum correlations. When exploring the role of the $r_{H}$ distance, it can be noticed that the correlation loss region becomes narrower along with the horizon size decrease. Simultaneously, the decline onset appears to be shifted slightly toward the horizon. This effect is best visible for the $D_{HH}=80$ (please refer to Fig. \ref{r2} (D)). To this end, the described behavior is obviously strongly related to the character of the local temperature. The corresponding density plots on the distance plane for the selected values of the $D_{HH}$ constant are given in Figs. \ref{r2} (E)-(F).

%%%%%%%%%%%%%%%%%%%%%%%%%%%%%%%%%%%%%%%%%%%%%%%%%%%%%%%%%%%%%%%%%%%%%%%%%%%%%%%%%%%%%%%%%%%%%%
\section{Discussion and conclusions}
\label{summary}
%%%%%%%%%%%%%%%%%%%%%%%%%%%%%%%%%%%%%%%%%%%%%%%%%%%%%%%%%%%%%%%%%%%%%%%%%%%%%%%%%%%%%%%%%%%%%%

The conducted analysis reveals that the quantum correlations, as captured within the MIN measure, allows to observe signatures of the black hole quantum atmosphere. In general, the obtained results suggest that the quantum correlations evolve continuously with the distance from the event horizon and that they are always finite in the Hartle-Hawking vacuum. This is to say, the whole atmosphere region is accessible at the quantum correlations level and its characteristics can be recovered within a correlation measure. In particular, two distinctive effects appear to be detectable in the atmosphere spectrum:

\begin{itemize}

\item[(i)]{The notable loss of correlations is observed at the macroscopic distance from the horizon ($r \approx 1.43 r_{H}$). It is important to note that this signature coincides with the Hawking radiation peak in the quantum atmosphere \cite{kim2017, eune2019}. Moreover, the correlation loss region is well-visible even when the horizon radius is varied, although it is shifted toward a black hole when its size is decreased. The latter is to some extent similar to the effect related to the influence of the black hole dimensionality on the atmosphere effective radius, as described in \cite{hod2016}.}

\item[(ii)]{The monotonic increase of quantum correlations is found to follow their initial degradation, as the distance from a black hole increases. This increase continues up to the maximum value, which is reached well outside the event horizon ($r > 4 r_{H}$). Note that the correlations are observed to saturate much faster with the distance than the local temperature itself, meaning that they stop exhibiting distinct characteristics of the atmosphere earlier. Sill, there is a strong interplay between both of these observables, suggesting potentially robust probing capabilities of a quantum correlation measures.}

\end{itemize}

Based on the described observations, it can be argued that the quantum correlations analysis constitutes promising venue to further discuss the Hawking radiation nature and to verify the Giddings argument from a new perspective \cite{giddings2016}. This can be done not only in theoretical terms but also has potential for an experimental realization.  Since the discussed phenomena are difficult or even impossible to examine in a real space-time, the analogue black holes may be viewed as an example to investigate the atmosphere concept with a help of quantum detectors \cite{barcelo2018, kolobov2021}. However, it means also that the consequences of the radiation origin location spans beyond the conventional black hole physics and profoundly impact behavior of quantum systems near the even horizon. In other words, the observed effects should be also considered in terms of the future research on information problems at large scales. Hence, the presented analysis supports at the same time the importance of the relativistic quantum information field in approaching problems such as the information paradox and other black hole phenomena measurable at the quantum level \cite{lanzagorta2014, hu2018}.

To this end, it is instructive to comment on the actual value of the $D_{HH}$ parameter. In principle, the physical meaning of this constant is largely unknown since its introduction \cite{eune2019}. However, the presented study shows that it may be possible to determine the $D_{HH}$ parameter based on the quantum correlation measure. This would simply require to associate the experimentally observed global correlation minimum (at $r/r_h\sim 1.43$) with the unknown parameter via the relations (\ref{eq7}) and (\ref{eq8}). Obviously, the main difficulty will be related to the construction of realistic setup capable of performing such measures as mentioned above.

\bibliography{bibliography}

\begin{thebibliography}{31}%
\makeatletter
\providecommand \@ifxundefined [1]{%
 \@ifx{#1\undefined}
}%
\providecommand \@ifnum [1]{%
 \ifnum #1\expandafter \@firstoftwo
 \else \expandafter \@secondoftwo
 \fi
}%
\providecommand \@ifx [1]{%
 \ifx #1\expandafter \@firstoftwo
 \else \expandafter \@secondoftwo
 \fi
}%
\providecommand \natexlab [1]{#1}%
\providecommand \enquote  [1]{``#1''}%
\providecommand \bibnamefont  [1]{#1}%
\providecommand \bibfnamefont [1]{#1}%
\providecommand \citenamefont [1]{#1}%
\providecommand \href@noop [0]{\@secondoftwo}%
\providecommand \href [0]{\begingroup \@sanitize@url \@href}%
\providecommand \@href[1]{\@@startlink{#1}\@@href}%
\providecommand \@@href[1]{\endgroup#1\@@endlink}%
\providecommand \@sanitize@url [0]{\catcode `\\12\catcode `\$12\catcode
  `\&12\catcode `\#12\catcode `\^12\catcode `\_12\catcode `\%12\relax}%
\providecommand \@@startlink[1]{}%
\providecommand \@@endlink[0]{}%
\providecommand \url  [0]{\begingroup\@sanitize@url \@url }%
\providecommand \@url [1]{\endgroup\@href {#1}{\urlprefix }}%
\providecommand \urlprefix  [0]{URL }%
\providecommand \Eprint [0]{\href }%
\providecommand \doibase [0]{https://doi.org/}%
\providecommand \selectlanguage [0]{\@gobble}%
\providecommand \bibinfo  [0]{\@secondoftwo}%
\providecommand \bibfield  [0]{\@secondoftwo}%
\providecommand \translation [1]{[#1]}%
\providecommand \BibitemOpen [0]{}%
\providecommand \bibitemStop [0]{}%
\providecommand \bibitemNoStop [0]{.\EOS\space}%
\providecommand \EOS [0]{\spacefactor3000\relax}%
\providecommand \BibitemShut  [1]{\csname bibitem#1\endcsname}%
\let\auto@bib@innerbib\@empty
%</preamble>
\bibitem [{\citenamefont {Hawking}(1975)}]{hawking1975}%
  \BibitemOpen
  \bibfield  {author} {\bibinfo {author} {\bibfnamefont {S.~W.}\ \bibnamefont
  {Hawking}},\ }\bibfield  {title} {\bibinfo {title} {Particle creation by
  black holes},\ }\href@noop {} {\bibfield  {journal} {\bibinfo  {journal}
  {Communications in Mathematical Physics}\ }\textbf {\bibinfo {volume} {43}},\
  \bibinfo {pages} {199} (\bibinfo {year} {1975})}\BibitemShut {NoStop}%
\bibitem [{\citenamefont {Hawking}(1976)}]{hawking1976}%
  \BibitemOpen
  \bibfield  {author} {\bibinfo {author} {\bibfnamefont {S.~W.}\ \bibnamefont
  {Hawking}},\ }\bibfield  {title} {\bibinfo {title} {Breakdown of
  predictability in gravitational collapse},\ }\href@noop {} {\bibfield
  {journal} {\bibinfo  {journal} {Physical Review D}\ }\textbf {\bibinfo
  {volume} {14}},\ \bibinfo {pages} {2460} (\bibinfo {year}
  {1976})}\BibitemShut {NoStop}%
\bibitem [{\citenamefont {Unruh}(1977)}]{unruh1977}%
  \BibitemOpen
  \bibfield  {author} {\bibinfo {author} {\bibfnamefont {W.~G.}\ \bibnamefont
  {Unruh}},\ }\bibfield  {title} {\bibinfo {title} {Origin of the particles in
  black-hole evaporation},\ }\href@noop {} {\bibfield  {journal} {\bibinfo
  {journal} {Physical Review D}\ }\textbf {\bibinfo {volume} {365}},\ \bibinfo
  {pages} {365} (\bibinfo {year} {1977})}\BibitemShut {NoStop}%
\bibitem [{\citenamefont {Giddings}(2016)}]{giddings2016}%
  \BibitemOpen
  \bibfield  {author} {\bibinfo {author} {\bibfnamefont {S.~B.}\ \bibnamefont
  {Giddings}},\ }\bibfield  {title} {\bibinfo {title} {Hawking radiation, the
  {S}tefan-{B}oltzmann law, and unitarization},\ }\href@noop {} {\bibfield
  {journal} {\bibinfo  {journal} {Physics Letters B}\ }\textbf {\bibinfo
  {volume} {754}},\ \bibinfo {pages} {39} (\bibinfo {year} {2016})}\BibitemShut
  {NoStop}%
\bibitem [{\citenamefont {Dey}\ \emph {et~al.}(2017)\citenamefont {Dey},
  \citenamefont {Liberati},\ and\ \citenamefont {Pranzetti}}]{dey2017}%
  \BibitemOpen
  \bibfield  {author} {\bibinfo {author} {\bibfnamefont {R.}~\bibnamefont
  {Dey}}, \bibinfo {author} {\bibfnamefont {S.}~\bibnamefont {Liberati}},\ and\
  \bibinfo {author} {\bibfnamefont {D.}~\bibnamefont {Pranzetti}},\ }\bibfield
  {title} {\bibinfo {title} {The black hole quantum atmosphere},\ }\href@noop
  {} {\bibfield  {journal} {\bibinfo  {journal} {Physics Letters B}\ }\textbf
  {\bibinfo {volume} {774}},\ \bibinfo {pages} {308} (\bibinfo {year}
  {2017})}\BibitemShut {NoStop}%
\bibitem [{\citenamefont {Dey}\ \emph {et~al.}(2019)\citenamefont {Dey},
  \citenamefont {Liberati}, \citenamefont {Mirzaiyan},\ and\ \citenamefont
  {Pranzetti}}]{dey2019}%
  \BibitemOpen
  \bibfield  {author} {\bibinfo {author} {\bibfnamefont {R.}~\bibnamefont
  {Dey}}, \bibinfo {author} {\bibfnamefont {S.}~\bibnamefont {Liberati}},
  \bibinfo {author} {\bibfnamefont {Z.}~\bibnamefont {Mirzaiyan}},\ and\
  \bibinfo {author} {\bibfnamefont {D.}~\bibnamefont {Pranzetti}},\ }\bibfield
  {title} {\bibinfo {title} {Black hole quantum atmosphere for freely falling
  observers},\ }\href@noop {} {\bibfield  {journal} {\bibinfo  {journal}
  {Physics Letters B}\ }\textbf {\bibinfo {volume} {797}},\ \bibinfo {pages}
  {134828} (\bibinfo {year} {2019})}\BibitemShut {NoStop}%
\bibitem [{\citenamefont {Ong}\ and\ \citenamefont {Good}(2020)}]{ong2020}%
  \BibitemOpen
  \bibfield  {author} {\bibinfo {author} {\bibfnamefont {Y.~C.}\ \bibnamefont
  {Ong}}\ and\ \bibinfo {author} {\bibfnamefont {M.~R.~R.}\ \bibnamefont
  {Good}},\ }\bibfield  {title} {\bibinfo {title} {{Quantum atmosphere of
  Reissner-Nordstr{\"o}m black holes}},\ }\href@noop {} {\bibfield  {journal}
  {\bibinfo  {journal} {Physical Review Research}\ }\textbf {\bibinfo {volume}
  {2}},\ \bibinfo {pages} {033322} (\bibinfo {year} {2020})}\BibitemShut
  {NoStop}%
\bibitem [{\citenamefont {Balbinot}\ and\ \citenamefont
  {Fabbri}(2022)}]{balbinot2022}%
  \BibitemOpen
  \bibfield  {author} {\bibinfo {author} {\bibfnamefont {R.}~\bibnamefont
  {Balbinot}}\ and\ \bibinfo {author} {\bibfnamefont {A.}~\bibnamefont
  {Fabbri}},\ }\bibfield  {title} {\bibinfo {title} {Quantum correlations
  across the horizon in acoustic and gravitational black holes},\ }\href@noop
  {} {\bibfield  {journal} {\bibinfo  {journal} {Physical Review D}\ }\textbf
  {\bibinfo {volume} {105}},\ \bibinfo {pages} {045010} (\bibinfo {year}
  {2022})}\BibitemShut {NoStop}%
\bibitem [{\citenamefont {Almheiri}\ \emph {et~al.}(2013)\citenamefont
  {Almheiri}, \citenamefont {Marolf}, \citenamefont {Polchinski},\ and\
  \citenamefont {Sully}}]{almheiri2013}%
  \BibitemOpen
  \bibfield  {author} {\bibinfo {author} {\bibfnamefont {A.}~\bibnamefont
  {Almheiri}}, \bibinfo {author} {\bibfnamefont {D.}~\bibnamefont {Marolf}},
  \bibinfo {author} {\bibfnamefont {J.}~\bibnamefont {Polchinski}},\ and\
  \bibinfo {author} {\bibfnamefont {J.}~\bibnamefont {Sully}},\ }\bibfield
  {title} {\bibinfo {title} {Black holes: complementarity or firewalls?},\
  }\href@noop {} {\bibfield  {journal} {\bibinfo  {journal} {Journal of High
  Energy Physics}\ }\textbf {\bibinfo {volume} {62}},\ \bibinfo {pages} {1}
  (\bibinfo {year} {2013})}\BibitemShut {NoStop}%
\bibitem [{\citenamefont {Kim}(2017)}]{kim2017}%
  \BibitemOpen
  \bibfield  {author} {\bibinfo {author} {\bibfnamefont {W.}~\bibnamefont
  {Kim}},\ }\bibfield  {title} {\bibinfo {title} {Origin of hawking radiation:
  firewall or atmosphere?},\ }\href@noop {} {\bibfield  {journal} {\bibinfo
  {journal} {General Relativity and Gravitation}\ }\textbf {\bibinfo {volume}
  {49}},\ \bibinfo {pages} {15} (\bibinfo {year} {2017})}\BibitemShut {NoStop}%
\bibitem [{\citenamefont {Eune}\ and\ \citenamefont {Kim}(2019)}]{eune2019}%
  \BibitemOpen
  \bibfield  {author} {\bibinfo {author} {\bibfnamefont {M.}~\bibnamefont
  {Eune}}\ and\ \bibinfo {author} {\bibfnamefont {W.}~\bibnamefont {Kim}},\
  }\bibfield  {title} {\bibinfo {title} {Test of quantum atmosphere in the
  dimensionally reduced {S}chwarzschild black hole},\ }\href@noop {} {\bibfield
   {journal} {\bibinfo  {journal} {Physics Letters B}\ }\textbf {\bibinfo
  {volume} {798}},\ \bibinfo {pages} {135020} (\bibinfo {year}
  {2019})}\BibitemShut {NoStop}%
\bibitem [{\citenamefont {Brunner}\ \emph {et~al.}(2014)\citenamefont
  {Brunner}, \citenamefont {Cavalcanti}, \citenamefont {Pironio}, \citenamefont
  {Stefano}, \citenamefont {Scarani}, \citenamefont {Valerio},\ and\
  \citenamefont {Wehner}}]{brunner2014}%
  \BibitemOpen
  \bibfield  {author} {\bibinfo {author} {\bibfnamefont {N.}~\bibnamefont
  {Brunner}}, \bibinfo {author} {\bibfnamefont {D.}~\bibnamefont {Cavalcanti}},
  \bibinfo {author} {\bibfnamefont {S.}~\bibnamefont {Pironio}}, \bibinfo
  {author} {\bibfnamefont {S.}~\bibnamefont {Stefano}}, \bibinfo {author}
  {\bibfnamefont {V.}~\bibnamefont {Scarani}}, \bibinfo {author} {\bibnamefont
  {Valerio}},\ and\ \bibinfo {author} {\bibfnamefont {S.}~\bibnamefont
  {Wehner}},\ }\bibfield  {title} {\bibinfo {title} {Bell nonlocality},\
  }\href@noop {} {\bibfield  {journal} {\bibinfo  {journal} {Reviews of Modern
  Physics}\ }\textbf {\bibinfo {volume} {86}},\ \bibinfo {pages} {419}
  (\bibinfo {year} {2014})}\BibitemShut {NoStop}%
\bibitem [{\citenamefont {He}\ \emph {et~al.}(2016)\citenamefont {He},
  \citenamefont {Xu},\ and\ \citenamefont {Ye}}]{he2016}%
  \BibitemOpen
  \bibfield  {author} {\bibinfo {author} {\bibfnamefont {J.}~\bibnamefont
  {He}}, \bibinfo {author} {\bibfnamefont {S.}~\bibnamefont {Xu}},\ and\
  \bibinfo {author} {\bibfnamefont {L.}~\bibnamefont {Ye}},\ }\bibfield
  {title} {\bibinfo {title} {{Measurement-induced-nonlocality for Dirac
  particles in Garfinkle–Horowitz–Strominger dilation space–time}},\
  }\href@noop {} {\bibfield  {journal} {\bibinfo  {journal} {Physics Letters
  B}\ }\textbf {\bibinfo {volume} {756}},\ \bibinfo {pages} {278} (\bibinfo
  {year} {2016})}\BibitemShut {NoStop}%
\bibitem [{\citenamefont {Kaczmarek}\ \emph {et~al.}(2023)\citenamefont
  {Kaczmarek}, \citenamefont {Szcz{\c e}{\'s}niak},\ and\ \citenamefont
  {Kais}}]{kaczmarek2023}%
  \BibitemOpen
  \bibfield  {author} {\bibinfo {author} {\bibfnamefont {A.~Z.}\ \bibnamefont
  {Kaczmarek}}, \bibinfo {author} {\bibfnamefont {D.}~\bibnamefont {Szcz{\c
  e}{\'s}niak}},\ and\ \bibinfo {author} {\bibfnamefont {S.}~\bibnamefont
  {Kais}},\ }\bibfield  {title} {\bibinfo {title} {Measurement-induced
  nonlocality for observers near a black hole},\ }\href@noop {} {\bibfield
  {journal} {\bibinfo  {journal} {Universe}\ }\textbf {\bibinfo {volume} {9}},\
  \bibinfo {pages} {199} (\bibinfo {year} {2023})}\BibitemShut {NoStop}%
\bibitem [{\citenamefont {Luo}\ and\ \citenamefont {Fu}(2011)}]{luo2011}%
  \BibitemOpen
  \bibfield  {author} {\bibinfo {author} {\bibfnamefont {S.}~\bibnamefont
  {Luo}}\ and\ \bibinfo {author} {\bibfnamefont {S.}~\bibnamefont {Fu}},\
  }\bibfield  {title} {\bibinfo {title} {Measurement-induced nonlocality},\
  }\href@noop {} {\bibfield  {journal} {\bibinfo  {journal} {Physical Review
  Letters}\ }\textbf {\bibinfo {volume} {106}},\ \bibinfo {pages} {120401}
  (\bibinfo {year} {2011})}\BibitemShut {NoStop}%
\bibitem [{\citenamefont {Hu}\ and\ \citenamefont {Fan}(2012)}]{hu2012}%
  \BibitemOpen
  \bibfield  {author} {\bibinfo {author} {\bibfnamefont {M.~L.}\ \bibnamefont
  {Hu}}\ and\ \bibinfo {author} {\bibfnamefont {H.}~\bibnamefont {Fan}},\
  }\bibfield  {title} {\bibinfo {title} {Dynamics of entropic
  measurement-induced nonlocality in structured reservoirs},\ }\href@noop {}
  {\bibfield  {journal} {\bibinfo  {journal} {Annals of Physics}\ }\textbf
  {\bibinfo {volume} {327}},\ \bibinfo {pages} {2343} (\bibinfo {year}
  {2012})}\BibitemShut {NoStop}%
\bibitem [{\citenamefont {Pan}\ and\ \citenamefont {Jing}(2008)}]{pan2008}%
  \BibitemOpen
  \bibfield  {author} {\bibinfo {author} {\bibfnamefont {Q.}~\bibnamefont
  {Pan}}\ and\ \bibinfo {author} {\bibfnamefont {J.}~\bibnamefont {Jing}},\
  }\bibfield  {title} {\bibinfo {title} {Hawking radiation, entanglement, and
  teleportation in the background of an asymptotically flat static black
  hole},\ }\href@noop {} {\bibfield  {journal} {\bibinfo  {journal} {Physical
  Review D}\ }\textbf {\bibinfo {volume} {78}},\ \bibinfo {pages} {065015}
  (\bibinfo {year} {2008})}\BibitemShut {NoStop}%
\bibitem [{\citenamefont {Farahi}\ and\ \citenamefont {{Pando
  Zayas}}(2014)}]{farahi2014}%
  \BibitemOpen
  \bibfield  {author} {\bibinfo {author} {\bibfnamefont {A.}~\bibnamefont
  {Farahi}}\ and\ \bibinfo {author} {\bibfnamefont {L.}~\bibnamefont {{Pando
  Zayas}}},\ }\bibfield  {title} {\bibinfo {title} {Gravitational collapse,
  chaos in {CFT} correlators and the information paradox},\ }\href@noop {}
  {\bibfield  {journal} {\bibinfo  {journal} {Physics Letters B}\ }\textbf
  {\bibinfo {volume} {734}},\ \bibinfo {pages} {31} (\bibinfo {year}
  {2014})}\BibitemShut {NoStop}%
\bibitem [{\citenamefont {Hu}\ \emph {et~al.}(2018)\citenamefont {Hu},
  \citenamefont {Hu}, \citenamefont {Wang}, \citenamefont {Peng}, \citenamefont
  {Zhang},\ and\ \citenamefont {Fan}}]{hu2018}%
  \BibitemOpen
  \bibfield  {author} {\bibinfo {author} {\bibfnamefont {M.~L.}\ \bibnamefont
  {Hu}}, \bibinfo {author} {\bibfnamefont {X.}~\bibnamefont {Hu}}, \bibinfo
  {author} {\bibfnamefont {J.}~\bibnamefont {Wang}}, \bibinfo {author}
  {\bibfnamefont {Y.}~\bibnamefont {Peng}}, \bibinfo {author} {\bibfnamefont
  {Y.~R.}\ \bibnamefont {Zhang}},\ and\ \bibinfo {author} {\bibfnamefont
  {H.}~\bibnamefont {Fan}},\ }\bibfield  {title} {\bibinfo {title} {Quantum
  coherence and geometric quantum discord},\ }\href@noop {} {\bibfield
  {journal} {\bibinfo  {journal} {Physics Reports}\ }\textbf {\bibinfo {volume}
  {762-764}},\ \bibinfo {pages} {1} (\bibinfo {year} {2018})}\BibitemShut
  {NoStop}%
\bibitem [{\citenamefont {Wu}\ \emph {et~al.}(2022)\citenamefont {Wu},
  \citenamefont {Cai}, \citenamefont {Peng},\ and\ \citenamefont
  {Zeng}}]{wu2022}%
  \BibitemOpen
  \bibfield  {author} {\bibinfo {author} {\bibfnamefont {S.~M.}\ \bibnamefont
  {Wu}}, \bibinfo {author} {\bibfnamefont {Y.~T.}\ \bibnamefont {Cai}},
  \bibinfo {author} {\bibfnamefont {W.~J.}\ \bibnamefont {Peng}},\ and\
  \bibinfo {author} {\bibfnamefont {H.~S.}\ \bibnamefont {Zeng}},\ }\bibfield
  {title} {\bibinfo {title} {Genuine {N}-partite entanglement and distributed
  relationships in the background of dilation black holes},\ }\href@noop {}
  {\bibfield  {journal} {\bibinfo  {journal} {The European Physical Journal C}\
  }\textbf {\bibinfo {volume} {82}},\ \bibinfo {pages} {412} (\bibinfo {year}
  {2022})}\BibitemShut {NoStop}%
\bibitem [{\citenamefont {Huang}\ \emph {et~al.}(2019)\citenamefont {Huang},
  \citenamefont {Yan}, \citenamefont {Wu},\ and\ \citenamefont
  {Hao}}]{huang2019}%
  \BibitemOpen
  \bibfield  {author} {\bibinfo {author} {\bibfnamefont {Y.}~\bibnamefont
  {Huang}}, \bibinfo {author} {\bibfnamefont {K.}~\bibnamefont {Yan}}, \bibinfo
  {author} {\bibfnamefont {Y.}~\bibnamefont {Wu}},\ and\ \bibinfo {author}
  {\bibfnamefont {X.}~\bibnamefont {Hao}},\ }\bibfield  {title} {\bibinfo
  {title} {Decoherence of quantum parameter estimation for open dirac particle
  in {Garfinkle–Horowitz–Strominger} dilation black hole},\ }\href@noop {}
  {\bibfield  {journal} {\bibinfo  {journal} {The European Physical Journal C}\
  }\textbf {\bibinfo {volume} {79}},\ \bibinfo {pages} {974} (\bibinfo {year}
  {2019})}\BibitemShut {NoStop}%
\bibitem [{\citenamefont {Li}\ \emph {et~al.}(2022)\citenamefont {Li},
  \citenamefont {Ming}, \citenamefont {Song}, \citenamefont {Ye},\ and\
  \citenamefont {Wang}}]{li2022}%
  \BibitemOpen
  \bibfield  {author} {\bibinfo {author} {\bibfnamefont {L.~J.}\ \bibnamefont
  {Li}}, \bibinfo {author} {\bibfnamefont {F.}~\bibnamefont {Ming}}, \bibinfo
  {author} {\bibfnamefont {X.~K.}\ \bibnamefont {Song}}, \bibinfo {author}
  {\bibfnamefont {L.}~\bibnamefont {Ye}},\ and\ \bibinfo {author}
  {\bibfnamefont {D.}~\bibnamefont {Wang}},\ }\bibfield  {title} {\bibinfo
  {title} {Quantumness and entropic uncertainty in curved space-time},\
  }\href@noop {} {\bibfield  {journal} {\bibinfo  {journal} {The European
  Physical Journal C}\ }\textbf {\bibinfo {volume} {82}},\ \bibinfo {pages}
  {726} (\bibinfo {year} {2022})}\BibitemShut {NoStop}%
\bibitem [{\citenamefont {Damour}\ and\ \citenamefont
  {Ruffini}(1976)}]{damour1976}%
  \BibitemOpen
  \bibfield  {author} {\bibinfo {author} {\bibfnamefont {T.}~\bibnamefont
  {Damour}}\ and\ \bibinfo {author} {\bibfnamefont {R.}~\bibnamefont
  {Ruffini}},\ }\bibfield  {title} {\bibinfo {title} {Black-hole evaporation in
  the {Klein-Sauter-Heisenberg-Euler} formalism},\ }\href@noop {} {\bibfield
  {journal} {\bibinfo  {journal} {Physical Review D}\ }\textbf {\bibinfo
  {volume} {14}},\ \bibinfo {pages} {332} (\bibinfo {year} {1976})}\BibitemShut
  {NoStop}%
\bibitem [{\citenamefont {Ge}\ and\ \citenamefont {Kim}(2008)}]{ge2008}%
  \BibitemOpen
  \bibfield  {author} {\bibinfo {author} {\bibfnamefont {X.~H.}\ \bibnamefont
  {Ge}}\ and\ \bibinfo {author} {\bibfnamefont {S.~P.}\ \bibnamefont {Kim}},\
  }\bibfield  {title} {\bibinfo {title} {Quantum entanglement and teleportation
  in higher dimensional black hole spacetimes},\ }\href@noop {} {\bibfield
  {journal} {\bibinfo  {journal} {Classical and Quantum Gravity}\ }\textbf
  {\bibinfo {volume} {25}},\ \bibinfo {pages} {075011} (\bibinfo {year}
  {2008})}\BibitemShut {NoStop}%
\bibitem [{\citenamefont {Lanzagorta}(2014)}]{lanzagorta2014}%
  \BibitemOpen
  \bibfield  {author} {\bibinfo {author} {\bibfnamefont {M.}~\bibnamefont
  {Lanzagorta}},\ }\href@noop {} {\emph {\bibinfo {title} {{Quantum Information
  in Gravitational Fields}}}}\ (\bibinfo  {publisher} {Morgan {\&} Claypool
  Publishers},\ \bibinfo {address} {San Rafael},\ \bibinfo {year}
  {2014})\BibitemShut {NoStop}%
\bibitem [{\citenamefont {Alsing}\ \emph {et~al.}(2006)\citenamefont {Alsing},
  \citenamefont {Fuentes-Schuller}, \citenamefont {Mann},\ and\ \citenamefont
  {Tessier}}]{alsing2006}%
  \BibitemOpen
  \bibfield  {author} {\bibinfo {author} {\bibfnamefont {P.~M.}\ \bibnamefont
  {Alsing}}, \bibinfo {author} {\bibfnamefont {I.}~\bibnamefont
  {Fuentes-Schuller}}, \bibinfo {author} {\bibfnamefont {R.~B.}\ \bibnamefont
  {Mann}},\ and\ \bibinfo {author} {\bibfnamefont {T.~E.}\ \bibnamefont
  {Tessier}},\ }\bibfield  {title} {\bibinfo {title} {Entanglement of {D}irac
  fields in noninertial frames},\ }\href@noop {} {\bibfield  {journal}
  {\bibinfo  {journal} {Physical Review A}\ }\textbf {\bibinfo {volume} {74}},\
  \bibinfo {pages} {032326} (\bibinfo {year} {2006})}\BibitemShut {NoStop}%
\bibitem [{\citenamefont {Harikrishnan}\ \emph {et~al.}(2022)\citenamefont
  {Harikrishnan}, \citenamefont {Jambulingam}, \citenamefont {Rohde},\ and\
  \citenamefont {Radhakrishnan}}]{harikrishnan2022}%
  \BibitemOpen
  \bibfield  {author} {\bibinfo {author} {\bibfnamefont {S.}~\bibnamefont
  {Harikrishnan}}, \bibinfo {author} {\bibfnamefont {S.}~\bibnamefont
  {Jambulingam}}, \bibinfo {author} {\bibfnamefont {P.~P.}\ \bibnamefont
  {Rohde}},\ and\ \bibinfo {author} {\bibfnamefont {C.}~\bibnamefont
  {Radhakrishnan}},\ }\bibfield  {title} {\bibinfo {title} {Accessible and
  inaccessible quantum coherence in relativistic quantum systems},\ }\href@noop
  {} {\bibfield  {journal} {\bibinfo  {journal} {Physical Review A}\ }\textbf
  {\bibinfo {volume} {105}},\ \bibinfo {pages} {052403} (\bibinfo {year}
  {2022})}\BibitemShut {NoStop}%
\bibitem [{\citenamefont {Tian}\ and\ \citenamefont {Jing}(2013)}]{tian2013}%
  \BibitemOpen
  \bibfield  {author} {\bibinfo {author} {\bibfnamefont {Z.}~\bibnamefont
  {Tian}}\ and\ \bibinfo {author} {\bibfnamefont {J.}~\bibnamefont {Jing}},\
  }\bibfield  {title} {\bibinfo {title} {{Measurement-induced-nonlocality via
  the Unruh effect}},\ }\href@noop {} {\bibfield  {journal} {\bibinfo
  {journal} {Annals of Physics}\ }\textbf {\bibinfo {volume} {333}},\ \bibinfo
  {pages} {76} (\bibinfo {year} {2013})}\BibitemShut {NoStop}%
\bibitem [{\citenamefont {Hod}(2016)}]{hod2016}%
  \BibitemOpen
  \bibfield  {author} {\bibinfo {author} {\bibfnamefont {S.}~\bibnamefont
  {Hod}},\ }\bibfield  {title} {\bibinfo {title} {Hawking radiation and the
  {Stefan–Boltzmann} law: The effective radius of the black-hole quantum
  atmosphere},\ }\href@noop {} {\bibfield  {journal} {\bibinfo  {journal}
  {Physics Letters B}\ }\textbf {\bibinfo {volume} {757}},\ \bibinfo {pages}
  {121} (\bibinfo {year} {2016})}\BibitemShut {NoStop}%
\bibitem [{\citenamefont {Barcel{\'o}}(2018)}]{barcelo2018}%
  \BibitemOpen
  \bibfield  {author} {\bibinfo {author} {\bibfnamefont {C.}~\bibnamefont
  {Barcel{\'o}}},\ }\bibfield  {title} {\bibinfo {title} {Analogue black-hole
  horizons},\ }\href@noop {} {\bibfield  {journal} {\bibinfo  {journal} {Nature
  Physics}\ }\textbf {\bibinfo {volume} {15}},\ \bibinfo {pages} {210}
  (\bibinfo {year} {2018})}\BibitemShut {NoStop}%
\bibitem [{\citenamefont {Kolobov}\ \emph {et~al.}(2021)\citenamefont
  {Kolobov}, \citenamefont {Golubkov}, \citenamefont {de~Nova},\ and\
  \citenamefont {Steinhauer}}]{kolobov2021}%
  \BibitemOpen
  \bibfield  {author} {\bibinfo {author} {\bibfnamefont {V.~I.}\ \bibnamefont
  {Kolobov}}, \bibinfo {author} {\bibfnamefont {K.}~\bibnamefont {Golubkov}},
  \bibinfo {author} {\bibfnamefont {J.~R.~M.}\ \bibnamefont {de~Nova}},\ and\
  \bibinfo {author} {\bibfnamefont {J.}~\bibnamefont {Steinhauer}},\ }\bibfield
   {title} {\bibinfo {title} {Observation of stationary spontaneous {Hawking}
  radiation and the time evolution of an analogue black hole},\ }\href@noop {}
  {\bibfield  {journal} {\bibinfo  {journal} {Nature Physics}\ }\textbf
  {\bibinfo {volume} {17}},\ \bibinfo {pages} {362} (\bibinfo {year}
  {2021})}\BibitemShut {NoStop}%
\end{thebibliography}%

\end{document}